# Spin-orbit-torque magnonics

V. E. Demidov,[1*] S. Urazhdin,[2] A. Anane,[3] V. Cros,[3] S. O. Demokritov[1]

[1]*Institute for Applied Physics and Center for Nanotechnology, University of Muenster, Corrensstrasse 2-4, 48149 Muenster, Germany*

[2]*Department of Physics, Emory University, Atlanta, GA 30322, USA*

[3]*Unité Mixte de Physique, CNRS, Thales, Université Paris-Saclay, 91767, Palaiseau, France*

The field of magnonics, which utilizes propagating spin waves for nano-scale transmission and processing of information, has been significantly advanced by the advent of the spin-orbit torque. The latter phenomenon can allow one to overcome two main drawbacks of magnonic devices – low energy efficiency of conversion of electrical signals into spin wave signals, and fast spatial decay of spin waves in thin-film waveguiding structures. At first glance, the excitation and amplification of spin waves by spin-orbit torques can seem to be straightforward. Recent research indicates, however, that the lack of the mode-selectivity in the interaction of spin currents with dynamic magnetic modes and the onset of dynamic nonlinear phenomena represent significant obstacles. Here, we discuss the possible route to overcoming these limitations, based on the suppression of nonlinear spin-wave interactions in magnetic systems with perpendicular magnetic anisotropy. We show that this approach enables efficient excitation of coherent magnetization dynamics and propagating spin waves in extended spatial regions, and is expected to enable practical implementation of complete compensation of spin-wave propagation losses.

**I. Introduction**

High-frequency spin waves propagating in thin ferromagnetic films have been utilized for the implementation of advanced microwave devices for many decades.[1-5] Unique features of these waves include electronic tunability by the magnetic field, very short (millimeter to submicrometer) wavelengths in the microwave frequency range, as well as controllability of propagation characteristics by the direction of the static magnetic field relative to the direction of wave propagation.[3-5] These features made spin waves very attractive for the implementation of a variety of devices for communication technologies, such as microwave filters, phase shifters, delay lines, multiplexers, etc. With the rapid increase in recent years of the operational frequencies of conventional electronic systems, in particular those for computation and information processing, high-frequency spin waves have become increasingly attractive for the implementation of integrated logic circuits and nanoscale wave-based computing systems.[6-8] This has motivated the emergence of a new research field - magnonics,[9-12] - which explores the possibility to utilize spin waves for transmission and processing of information at nanoscale as an alternative to conventional CMOS-based electronic circuits.

Despite numerous advantages provided by spin waves, magnonic devices currently suffer from two drawbacks. First, the inductive mechanism traditionally utilized to convert electrical signals into spin waves and back, is characterized by relatively low conversion efficiency.[13-18] Even if low conversion losses can be achieved in traditional macroscopic spin-wave devices with millimeter-scale dimensions, miniaturization of magnonic devices down to the sub-micrometer scale inevitably results in large conversion losses unacceptable for real-world applications. The second drawback of microscopic magnonic devices is associated with large propagation losses of spin waves in nanometer-thick magnetic films. In macroscopic spin-wave devices based on high-quality micrometer-thick films of Yttrium Iron Garnet (YIG), the decay length of spin waves reaches several millimeters. However, in microscopic waveguides patterned from ultrathin YIG films, the decay length typically does not exceed a few tens on micrometers[19-21].

To make the microscopic magnonic devices technologically competitive, it is necessary to overcome their two main drawbacks described above. This may become possible thanks to the advent of spin-torque phenomena[22,23], which allow reduction of magnetic damping, and even its complete compensation, enabling excitation of high-frequency magnetization oscillations by dc electrical currents[24,25]. In particular, it has been shown that spin torques created by spin-polarized electric currents or by non-local spin injection can excite propagating spin waves[26-32], and that this excitation mechanism is intrinsically fast[31] and highly power-efficient.[30] These achievements clearly demonstrated a large potential of spin torques for magnonics. However, the devices proposed in Refs. 26-32 utilized current injection through a magnetic nanocontact, resulting only in local spin torque. It could not provide compensation of damping over spatially extended regions, which is of vital importance for magnonic applications that require enhanced spin wave propagation.

In contrast to the conventional spin torques produced by the local current injection through conducting magnetic materials, the spin-orbit torque (SOT) associated with charge to spin conversion in the bulk or at interfaces in material systems with strong spin-orbit interaction,[33-36] provides the ability to control magnetic damping in spatially extended regions of both conducting and insulating magnetic materials[37-39]. This may enable decay-free propagation of spin waves, or even their true amplification in magnetic nanostructures. In recent years, this possibility was intensively explored for macroscopic[40-43] and microscopic[44-48] magnonic devices. The largest effect reported so far was achieved in microscopic magnonic waveguides based on ultrathin YIG films,[46] where a nearly tenfold increase in the propagation length of coherent spin waves was demonstrated. The high efficiency of the system studied in Ref. 46 permitted experimental access to the operational regime close and even above the point at which damping is expected to be completely compensated by SOT. Contrary to naive expectations, it was found that the onset of nonlinear damping in the overdriven magnon system does not allow one to achieve complete compensation of spin-wave propagation losses, required for true amplification of coherent spin waves by SOT.

Note that the same fundamental mechanisms also prevent SOT-induced excitation of coherent magnetization oscillations in spatially extended systems.[39,49] In Ref. 50, it was shown that the adverse effects of nonlinear damping can be suppressed by localizing the injection of spin current to a nanoscale region of the magnetic system. This approach enabled the implementation of SOT-driven magnetic nano-oscillators that exhibit coherent high-frequency magnetization oscillations[50-61]. However, it also imposed strict limitations on the layout of these devices, hindering their utilization as sources of propagating spin waves. As a result, it took a significant time to understand how confined-area SOT-driven oscillators can be integrated with a medium supporting spin-wave propagation.[62,63] We emphasize that, while spin-wave excitation by SOT was demonstrated in Refs. 62,63, these works have not resolved the general problem of the adverse effects of nonlinear damping.

In this brief perspective article, we present our vision of the main outstanding issues hindering further developments in the field of SOT-driven magnonics, and outline an approach that can help to overcome them. We first discuss the key recent experimental results revealing the physical nature of these issues. In particular, we show that the functionality of SOT magnonic devices is strongly limited by the lack of spin-wave mode selectivity of the interaction of spin currents with the magnetization, which results in the simultaneous enhancement by SOT of many incoherent spin-wave modes. This enhancement causes an onset of the nonlinear scattering of coherent spin waves, which counteracts their enhancement by the anti-damping effect of SOT. We discuss a recently proposed approach that allows efficient control of nonlinear damping, by utilizing reduction of ellipticity of magnetization precession in magnetic films where the dipolar anisotropy is compensated by the interfacial perpendicular magnetic anisotropy (PMA). This approach has been recently shown to enable SOT-driven excitation of coherent magnetization auto-oscillations in spatially extended systems based on conductive[64] and insulating[65] magnetic materials. It was also demonstrated to enable efficient excitation of coherent propagating spin waves in magnetic insulators.[65] We project that this approach can also enable the implementation of decay-free propagation of spin waves, resolving, in this way, the main issues limiting the progress in nano-magnonics.

## II. Amplification of spin waves by SOT

Figure 1 presents the results of the experiment that demonstrated highly efficient compensation of propagation losses of spin waves by SOT[46]. The test devices used in this study (Fig. 1(a)) are 1 μm-wide spin-wave waveguides patterned from a 20 nm-thick YIG film covered by an 8 nm-thick layer of Pt. Propagating spin waves are excited in the waveguide by microwave current flowing through a 3 μm wide and 250 nm thick Au inductive antenna. Additionally, a dc electrical current $I$ is applied through the Pt layer of the waveguide. The electrical current is converted into an out-of-plane pure spin current $I_S$ (see the inset in Fig. 1(a)), due to the spin-Hall effect (SHE)[33-36] in Pt. In turn, the spin current exerts an anti-damping spin-transfer torque[37] on the magnetization $M$ in YIG, which is expected to compensate the propagation losses of the coherent spin wave emitted by the antenna.

The effects of SOT on the spin-wave propagation were studied by using micro-focus Brillouin light scattering (BLS) spectroscopy.[66] The probing laser light was focused through the sample substrate into a diffraction-limited spot on the YIG/Pt film (Fig. 1(a)), and the modulation of light due to its interaction with magnetic excitations in YIG was analyzed. The resulting signal – the BLS intensity – is proportional to the intensity of magnetic oscillations at the position of the probing spot, which enables mapping of the intensity of propagating spin waves with high spatial resolution (Fig. 1(b)).

Figure 1(c) shows several representative dependences of spin-wave intensity on the propagation coordinate $x$, obtained at different dc currents in the Pt layer. As seen from these data, spin waves in the waveguide experience well-defined exponential decay $\sim \exp(-2x/\xi)$, where $\xi$ is the decay length, defined as the distance over which the wave amplitude decreases by a factor of e. The decay length increases with the increase of the dc current, as expected for the effects of SOT on the effective magnetic damping. Figure 1(d) shows the current dependences of the decay length $\xi$ and of its inverse value – the decay constant, which is proportional to the effective Gilbert damping constant. The decay length monotonically increases at small currents, reaches a maximum at a certain current $I = I_C$, and then abruptly decreases at larger currents. Independent measurements were used to identify $I_C$ as the critical current, at which SOT is expected to completely compensate the natural damping in YIG. Thus, one could expect that $\xi$ diverges at $I = I_C$ (the decay constant vanishes), and that the propagating wave becomes spatially amplified at $I > I_C$. These naive expectations are clearly inconsistent with the experimental findings.

This experiment demonstrated that SOT is capable to increasing the propagation length of spin waves by nearly an order of magnitude. For the studied system, this corresponded to the increase of the spin-wave intensity at the output of a 10 μm-long transmission line by three orders of magnitude. Simultaneously, this experiment showed that the simple picture of SOT effects on the damping of coherent propagating spin waves becomes invalid in the vicinity of the point of the complete damping compensation. As will be discussed below, in the regime of large currents, it is necessary to take into account not only that SOT reduces the effective damping, but also that it strongly enhances incoherent magnetic fluctuations. The nonlinear scattering of coherent waves from these fluctuations represents an additional damping channel counteracting the anti-damping effect of SOT.

### III. Excitation of spin waves by SOT

While SOT-induced coherent magnetic auto-oscillations have been achieved in several nanomagnetic device geometries[50-53], and many other novel SOT oscillators have been proposed in recent years[54-61], none of them provided the possibility to generate coherent propagating spin waves. This is mainly associated with the limitations imposed by the device layout necessary to achieve coherent auto-oscillations. Indeed, as was shown in Refs. 39,49, coherent oscillations cannot be excited in spatially extended systems with arbitrary shape, due to the nonlinear interactions with incoherent spin-wave modes strongly enhanced by SOT. To suppress these nonlinear effects, spin current must be injected into a nanoscale region of the magnetic system, which imposes strict limitations on the layout of SOT-driven auto-oscillators.

After an intensive search for a suitable geometry, in Ref. 62, we proposed a new concept of nano-notch SOT auto-oscillators directly incorporated into a magnonic nano-waveguide. These devices (Fig. 2(a)) were based on 180 nm-wide nano-waveguides patterned from a Permalloy(Py)(15 nm)/Pt(4 nm) bilayer. Ion milling was used to pattern a rectangular 200 nm-wide and 10 nm-deep notch in the top Py layer of the waveguide, forming a nano-oscillator that serves as the spin-wave source. When electric current $I$ flows through the device, SHE in Pt injects pure spin current $I_S$ into the Py layer, producing anti-damping SOT acting on its magnetization $M$. The thickness-averaged magnitude of the anti-damping torque is inversely proportional to the thickness of the magnetic layer. Thus, the SOT effects on the 5 nm-thick Py layer in the nano-notch area are significantly larger than on the 15 nm-thick Py waveguide. As the current $I$ is increased, damping becomes completely compensated in the nano-notch region, resulting in the local excitation of magnetization auto-oscillations.

These devices exhibit two-mode auto-oscillations (Fig. 2(b)). The frequency of the low-frequency mode, excited at small currents, is smaller than the frequencies of propagating spin waves in the 15 nm-thick waveguide. Correspondingly, the spatial mapping of the oscillations by micro-focus BLS showed that this mode is localized in the nano-notch and does not emit spin waves into the waveguide (Fig. 3(c)). In contrast, the high-frequency mode, excited at larger currents, was found to efficiently emit propagating spin waves (Fig. 3(d)). This emission was found to be strongly unidirectional, with the preferential direction controlled by the direction of the static magnetic field. Additionally, it was shown that the propagation length of emitted spin waves is enhanced by non-zero spin current injected over the entire length of the waveguide, by up to a factor of three.

The system proposed in Ref. 62 combines all the advantages provided by SOT to locally excite propagating spin waves, and to simultaneously enhance their propagation characteristics. The achieved enhancement can be further increased by the material engineering and geometry optimization. Additionally, the proposed approach can be easily scaled to chains of SOT nano-oscillators coupled via propagating spin waves, facilitating the development of novel nanoscale signal processing circuits such as logic and neuromorphic computing networks. We would like to note, however, that in spite of all these advantages, the proposed system also suffers from the limitations associated with the nonlinear spin-wave scattering. The latter adversely affects the oscillation characteristics of nano-notch devices at large driving currents, resulting in a strong reduction of the intensity of emitted spin waves (Fig. 2(b)), and does not allow one to achieve complete compensation of their propagation losses.

### IV. Control of nonlinear damping in SOT-driven devices

The effects of SOT on the magnetization are often approximated as a simple modification of the effective magnetic damping[37]. This simple picture neglects the effects of fluctuations always present at finite temperatures. Analysis taking into account this contribution reveals another important effect of SOT – namely, it drives the spin system out of thermal equilibrium, resulting in the enhancement of magnetic fluctuations, which can equivalently be described as excitation of a large number of incoherent spin waves (magnons) spread over a broad interval of frequencies and wavelengths.[39,67] The importance of this contribution has been recognized starting with the first experiments on the interaction of spin currents with magnetization.[49] Later on, excitation of incoherent magnons by SOT and their propagation received a significant attention as a mechanism for transmission of spin information in magnetic insulators.[68-72] However, the spectral characteristics of magnons excited by SOT remained unknown for a long time. This issue was addressed experimentally in Ref. 67. In this work, by using the BLS technique, we were able to study the effects of SOT on the magnon distribution in a Py/Pt bilayer over a significant spectral range.

Figure 3(a) shows the BLS spectra reflecting the spectral density of magnons recorded with and without dc current in Pt. The spectrum obtained at $I = 0$ characterizes the magnons present in Py due to thermal fluctuations at room temperature. When the current is applied, the populations of all magnon states significantly increase due to the effects of SOT. Note that the enhancement of the magnon population is most significant at low frequencies, and rapidly decreases with the increase of the frequency of magnons (Fig. 3(b)), which can be associated with the larger relaxation rates of magnons with higher frequencies. With the increase in the SOT strength, the dominant mode at the lowest frequency $f_{min}$ is expected to transition to the auto-oscillation regime at its damping compensation point. However, the increase in the amplitudes of all the other modes leads to their nonlinear coupling, which results in the energy flow from the dominant mode (see, e.g., results of micromagnetic simulations Fig. 5d in Ref. 64). This process represents an onset of additional (nonlinear) damping, which prevents complete damping compensation by SOT.

The adverse effects of nonlinear damping can be reduced by suppressing the amplitudes of parasitic incoherent modes. This approach was used to achieve SOT-driven coherent magnetization oscillations by utilizing local injection of spin current.[39,50] In this geometry, parasitic incoherent spin waves quickly escape from the localized active area, resulting in reduced nonlinear damping of the dominant mode. This approach generally restricts the active region to nanoscale dimensions, limiting the achievable dynamical coherence and the possibilities for the magnonic device integration. Moreover, this approach cannot be extended to the compensation of propagation losses of spin waves by SOT over extended spatial areas.

We have recently proposed a new, more efficient approach to suppression of nonlinear damping, based on the direct control of the mode coupling mechanisms.[64] In this work, we showed that the nonlinear spin wave coupling is predominantly determined by the ellipticity of magnetization precession, which is controlled in thin magnetic films by the demagnetizing effects. We demonstrated that, by using magnetic films with suitably tailored perpendicular magnetic anisotropy (PMA), one can compensate the dipolar anisotropy and achieve almost circular precession, resulting in suppression of nonlinear damping. As a result, we were able to achieve complete damping compensation and excitation of coherent magnetization auto-oscillations by SOT, in a simple system with uniform spatially-extended injection of spin current.

Figure 4(a) shows the layout of the test device studied in this work. It consists of a 5 nm-thick magnetic disk fabricated on top of an 8 nm-thick Pt strip, which plays the role of a spin-current injector. We emphasize, that it is well known from early studies[39,49], that in this geometry it is impossible to achieve complete damping compensation and excitation of magnetization auto-oscillations in magnetic systems with in-plane anisotropy, due to the adverse effects of the nonlinear damping. Indeed, no transition to auto-oscillations is observed when Permalloy is used as the magnetic disk material (Fig. 4(b)). When the current $I$ is applied through the Pt strip, the recorded BLS spectra show an enhancement of magnetic fluctuations in Py. The intensity of fluctuations gradually increases with increasing $I$ at currents smaller than the critical value $I_C$. However, at $I > I_C$, the intensity of fluctuations saturates, while their spectral width significantly increases.

The observed behaviors change dramatically, if the magnetic disk is made from a Co/Ni bilayer, with the relative thicknesses of Co and Ni adjusted so that the effective PMA field $H_a$ compensates the saturation magnetization $4\pi M$ of the bilayer (Fig. 4(c)). The compensation of the dipolar anisotropy by PMA results in nearly circular magnetization precession trajectory. This can be contrasted with the magnetization precession of the Py disk, which exhibits a large ellipticity (see the insets in Figs. 4(b) and 4(c)). As seen from Fig. 4(c), at $I < I_C$, the effects of SOT in CoNi disk are very similar to those in the Py disk. However, at $I > I_C$, a narrow intense spectral peak emerges for CoNi, marking a transition to the auto-oscillation regime. This result shows that the nonlinear damping that prevents the onset of auto-oscillations in the Py disk is suppressed in CoNi.

The experimental results described above, together with the micromagnetic simulations of these systems, clearly demonstrated a route for overcoming the limitations imposed by the nonlinear damping, by utilizing PMA materials with tailored anisotropy strength. This approach allows one to achieve complete compensation of the magnetic damping, and excitation of coherent magnetization auto-oscillations by SOT, without confining the spin-current injection to a nanoscale area. This can enable the implementation of SOT-driven oscillators with spatially extended active area, capable of generating microwave signals with technologically relevant power levels and coherence, and thus circumventing the challenges of phase locking of a large number of oscillators with nano-scale dimensions. The proposed approach also provides a route for the implementation of spatially extended amplification of coherent propagating spin waves.

**V. Highly efficient SOT-driven spin-wave emission in YIG films with PMA**

The main challenges in the application of the approach described above to metallic multilayers with PMA are associated with their large magnetic damping, and the strong spatial inhomogeneity of magnetic properties typical for these systems. Recent progress in the deposition of high-quality nanometer-thick films of YIG can allow one to overcome these limitations. In particular, it was shown in Ref. 73 that Bi doping of ultrathin YIG films can facilitate a large PMA controllable by the epitaxial strain and growth-induced anisotropies, while preserving their low-damping characteristics. These properties make Bi-doped YIG films uniquely suitable for SOT-based magnonic devices utilizing suppression of nonlinear damping by the anisotropy compensation.

Figure 5(a) shows the layout of the SOT device based on an extended 20-nm thick film of Bi-doped YIG (BiYIG) ($Bi_1Y_2Fe_5O_{12}$) with PMA.[65] To achieve suppression of the nonlinear interactions, the strength of PMA is tuned to exactly compensate the dipolar anisotropy. The device utilizes a simple large-scale injector of spin current, formed by a 6-nm thick Pt strip line with the width of 1 μm and the length of 4 μm. Similar simple spin-current injectors were utilized in many prior experiments on YIG films without PMA.[68-72] We emphasize that it was well established in those studies, that SOT-driven emission of coherent propagating SW cannot be achieved in in-plane magnetized YIG, even at large driving currents exceeding $I_C$. Instead, the dominant mode saturates in the vicinity of the point of complete damping compensation, similar to the Py disks on Pt (Fig. 4(b)).[72]

In contrast, the devices based on BiYIG were found to exhibit a clear transition to coherent auto-oscillations at moderate current densities in the Pt strip. By using local detection of the auto-oscillation spectra with BLS (Fig. 5(b)), we confirmed the high spatial coherence of the auto-oscillations, whose frequency remains unchanged over the entire active area above the 1×4 μm Pt injector. In contrast to most of the previously demonstrated SOT oscillators, the intensity of the auto-oscillation peak monotonically increases with the increase of the current strength over a broad range of currents (Fig. 5(c)), consistent with suppression of nonlinear interactions that typically result in the degradation of the oscillation characteristics at large current densities (see, e.g., Fig. 2(b)).

Another consequence of the precise compensation of the dipolar anisotropy by PMA is the absence of the nonlinear frequency shift of the auto-oscillations with the increase of their intensity (Fig. 5(c)). The spectral coherence of auto-oscillations is known to be significantly enhanced when the nonlinear shift is small, as it hence reduces the coupling between the oscillation amplitude and its phase.[74] In addition, the nonlinear frequency shift, common for in-plane magnetized devices (see, e.g., Fig. 2(b)), is well-known to result in self-localization of auto-oscillation into a nonlinear spin-wave bullet.[75,76] In the absence of the nonlinear shift, the self-localization is not expected to occur, enabling emission of coherent spin waves into the surrounding film. Indeed, spatially resolved BLS measurements (Fig. 5(d)) confirm efficient spin-wave emission from the active device area. As seen from these data, the oscillations excited due to SOT are not localized in the area above the Pt line, but significantly extend into the surrounding BiYIG film.

Analysis performed in Ref. 65 showed that the wavelength of the emitted spin waves is about 300 nm, as determined by the difference in the PMA strength in the bare BiYIG film and in the BiYIG/Pt bilayer. The wavelength can be controlled by varying the difference between these anisotropies. Additionally, the frequency of auto-oscillations and the wavelength of the emitted spin waves can be controlled by the current, in devices where the strength of PMA is increased beyond the point of exact compensation of the dipolar anisotropy. These conditions correspond to the positive nonlinear frequency shift, which allows one to increase the auto-oscillation frequency and decrease the wavelength of emitted spin waves by the increase of the current in the Pt line.

The results described above clearly demonstrate that suppression of the adverse nonlinear damping in materials with PMA allows one to overcome the most significant limitations hindering the practical realization of advantages provided by SOT in magnonics. We believe that these findings should spur further progress in this field.

**VI. Conclusions**

Downscaling of magnonic devices presents a large number of new challenges, which must be addressed before spin-wave technology becomes a competitive alternative to conventional CMOS-based microelectronics. Simultaneously, downscaling provides novel opportunities unavailable in traditional macroscopic-scale spin-wave devices. Spin-orbit torque is one of the striking examples of physical mechanisms that become particularly efficient at nanoscale. Utilization of such phenomena is necessary for the successful development of nano-magnonics, enabling the implementation of nanodevices that are more efficient and more attractive for real-world applications than their previously demonstrated macroscopic counterparts. Naturally, novel physical phenomena, which have not yet been fully understood, require intense research, often revealing new unexpected difficulties and challenges as well as new opportunities. This is clearly the case with SOT-driven excitation and amplification of spin waves. The results obtained in early experiments in the "small-amplitude" regime could be easily interpreted by using a simple interpretation of the effects of SOT in terms of the variation of the effective damping. However, with the development of highly-efficient SOT systems, where complete damping compensation can be achieved, new challenges have emerged, which can only be overcome based on the deep insight into nonlinear dynamic processes in strongly-driven spin systems. We believe that the recently proposed approach discussed in this article will finally enable efficient integration of spin-orbitronics and magnonics, and will accelerate new developments in the latter field.

**ACKNOWLEDGMENTS**

This work was supported in part by the Deutsche Forschungsgemeinschaft (Project No. 423113162), the NSF Grant No. ECCS-1804198, by the French ANR Grants 15-CE08-0030-01 (ISOLYIG) and the ANR-18-CE24-0021 (MAESTRO). The authors acknowledge the essential contribution of the PhD students that have contributed to some of the studies presented in this article: Martin Collet, Boris Divinskiy, Michael Evelt, Diane Gouéré, Hugo Merbouche, and Lucile Soumah.

**Data Availability**

The data that support the findings of this study are available from the corresponding author upon reasonable request.

x

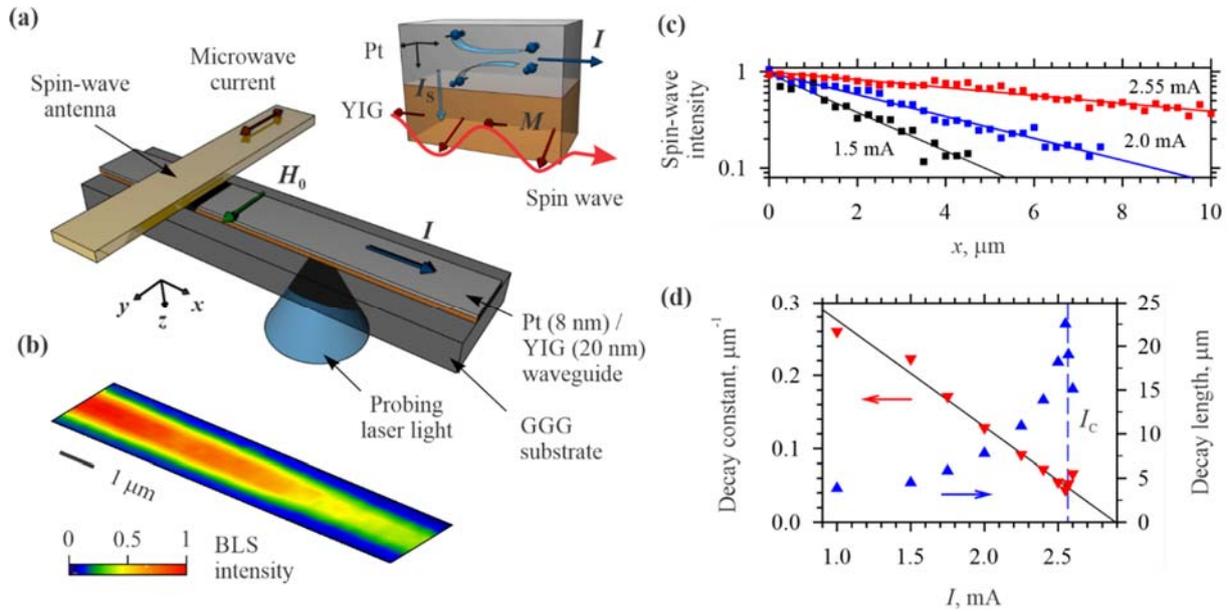

Fig. 1. (a) Schematic of the SOT device based on the Pt/YIG bilayer waveguide, and the experimental setup. (b) Measured spatial map of the BLS intensity, which is proportional to the local spin-wave intensity. (c) Dependences of the spin-wave intensity on the propagation coordinate, recorded at different dc currents in the Pt layer, as labeled. Symbols – experimental data, curves – exponential fits. (d) Current dependences of the decay length (point-up triangles) and of its inverse value (point-down triangles). $I_C$ marks the critical current, at which SOT is expected to completely compensate the natural damping. Reproduced from M. Evelt, V. E. Demidov, V. Bessonov, S. O. Demokritov, J. L. Prieto, M. Muñoz, J. Ben Youssef, V. V. Naletov, G. de Loubens, O. Klein, M. Collet, K. Garcia-Hernandez, P. Bortolotti, V. Cros and A. Anane, Appl. Phys. Lett. **108**, 172406 (2016), with the permission of AIP Publishing.

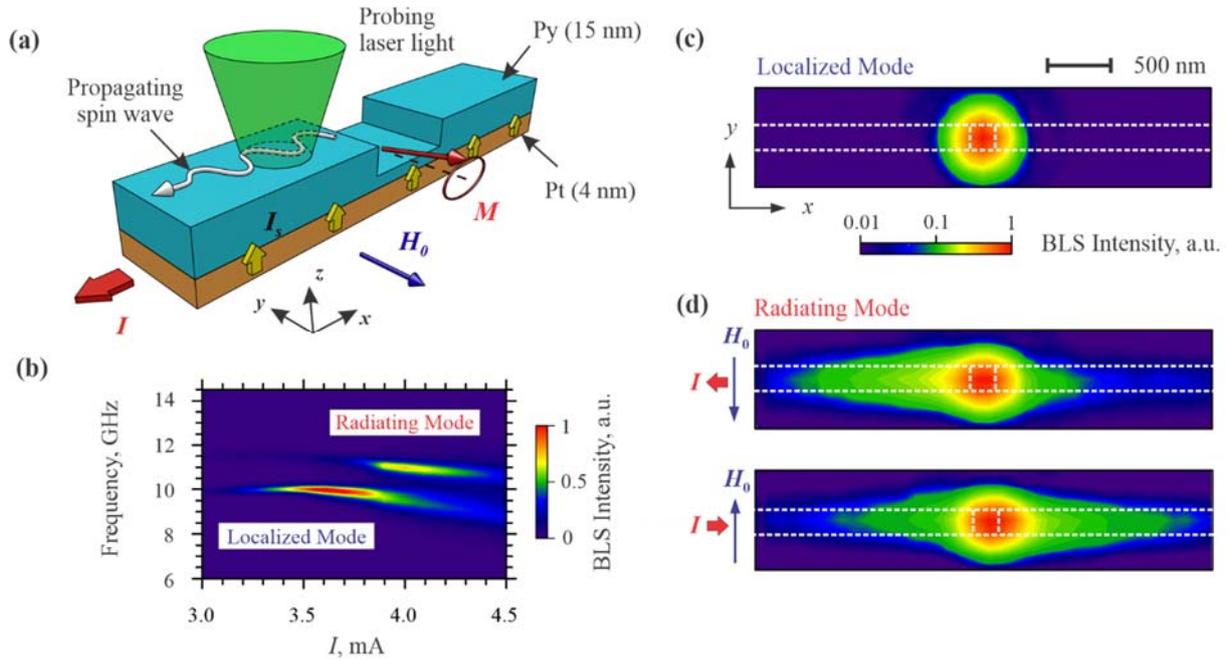

Fig. 2. (a) Schematic of the SOT device based on the notched Pt/Py bilayer nano-waveguide, and the experimental setup. (b) Normalized color-coded map of the BLS intensity in the frequency-current coordinates. (c) and (d) Color-coded spatial maps of the BLS intensity, measured at the frequencies of the localized and of the radiating modes, as labeled. Dashed lines on the maps show the outlines of the waveguide and of the nano-notch. Reproduced with permission from B. Divinskiy, V. E. Demidov, S. Urazhdin, R. Freeman, A. B. Rinkevich, and S. O. Demokritov, Adv. Mater. **30**, 1802837 (2018). Copyright 2018 John Wiley and Sons.

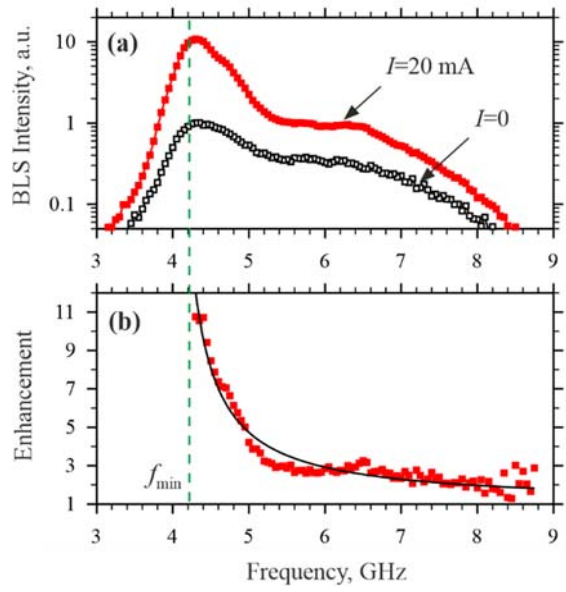

Fig. 3. Effects of SOT on spectral magnon distribution for an SOT device based on crossed Pt and Py strips. (a) BLS spectra, reflecting the spectral density of magnons, recorded at $I$=0 and 20 mA, as labeled. Vertical dashed line marks the frequency of the lowest-energy magnon state. (b) Enhancement of the magnon population at $I$=20 mA. Reproduced from V. E. Demidov, S. Urazhdin, B. Divinskiy, V. D. Bessonov, A. B. Rinkevich, V.V. Ustinov, and S. O. Demokritov, Nat. Commun. **8**, 1579 (2017); licensed under a Creative Commons Attribution (CC BY) license.

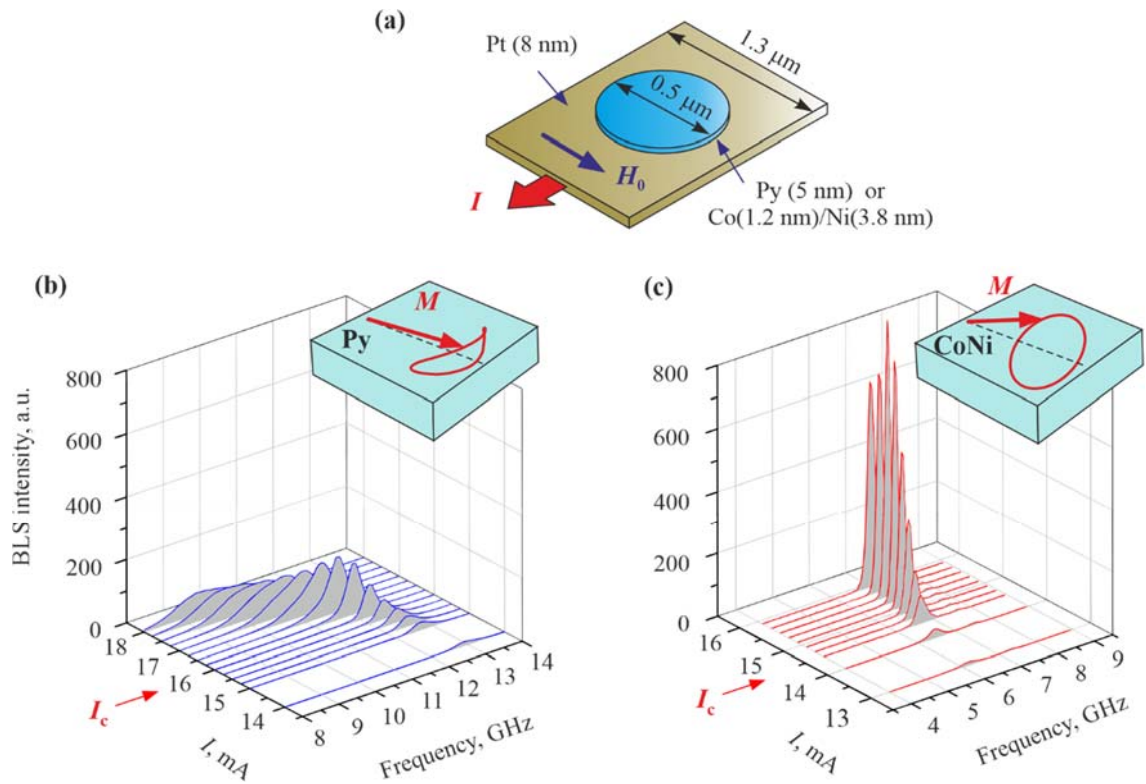

Fig. 4. (a) Layout of the test SOT devices based on Py and CoNi disks on the Pt strip. (b) and (c) BLS spectra of magnetic oscillations *vs* current for Py and CoNi disks, respectively. $I_C$ marks the critical current, at which SOT is expected to completely compensate the natural linear magnetic damping. Insets illustrate the ellipticities of magnetization precession in Py and CoNi, respectively. Reproduced from B. Divinskiy, S. Urazhdin, S. O. Demokritov, and V. E. Demidov, Nat. Commun. **10**, 5211 (2019); licensed under a Creative Commons Attribution (CC BY) license.

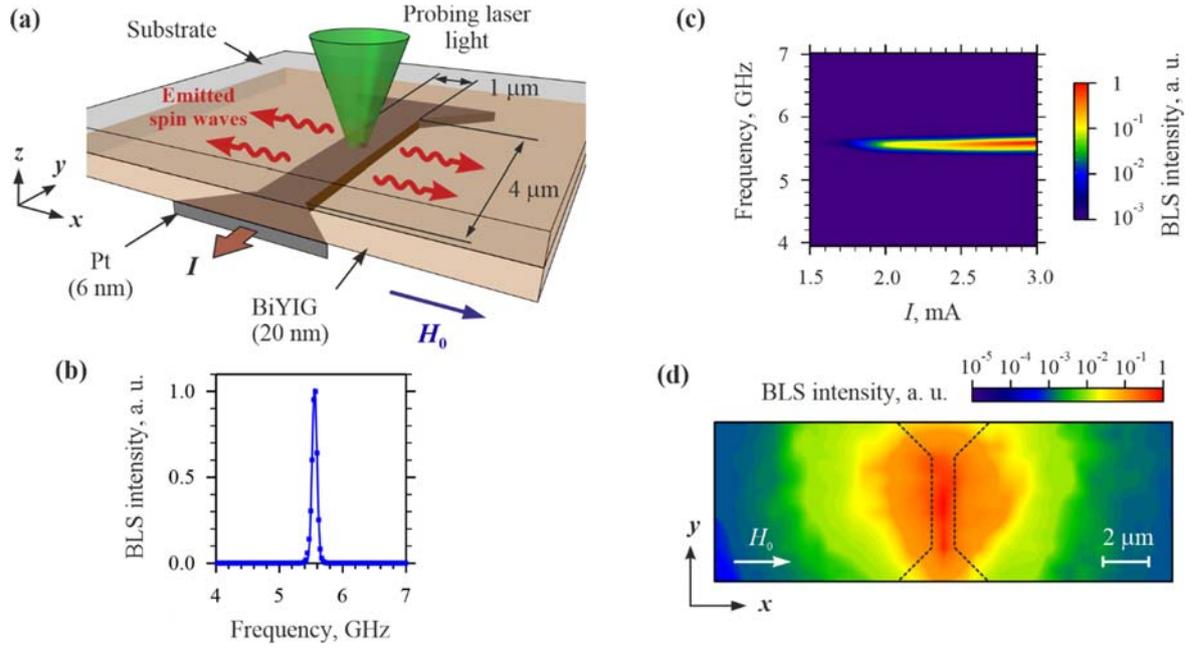

Fig. 5. (a) Schematic of the SOT devices utilizing Pt strip on extended BiYIG film. (b) Representative BLS spectrum of magnetization auto-oscillations recorded at $I$=2.5 mA. Note that the width of the measured spectral peak is determined by the limited frequency resolution of BLS. (c) Color-coded BLS intensity in the current-frequency coordinates. (d) Color-coded spatial map of the BLS intensity recorded by rastering the probing laser spot over 20 μm by 7 μm area. Dashed lines show the contours of the Pt line. Reproduced with permission from M. Evelt, L. Soumah, A. B. Rinkevich, S. O. Demokritov, A. Anane, V. Cros, J. Ben Youssef, G. de Loubens, O. Klein, P. Bortolotti, and V. E. Demidov, Phys. Rev. Appl. **10**, 041002 (2018). Copyright 2018 by the American Physical Society.